\documentclass[a4paper]{article}

\usepackage{INTERSPEECH2021}
\usepackage[table]{xcolor}
\usepackage{hyperref}
\usepackage{graphicx}
\hypersetup{colorlinks=true}
\usepackage{multirow}
\usepackage{float}
\usepackage{array}
\usepackage{enumitem}
\usepackage{soul}
\newcolumntype{C}[1]{>{\centering\arraybackslash}m{#1}}

\title{Personalized Keyphrase Detection using \\ Speaker and Environment Information}
\name{Rajeev Rikhye$^{*}$, Quan Wang$^{*}$, Qiao Liang, Yanzhang He, \\
Ding Zhao, Yiteng (Arden) Huang, Arun Narayanan, Ian McGraw\thanks{* Equal contribution. }}
\address{
Google LLC
}
\email{
\{\href{mailto:rvrikhye@google.com}{rvrikhye},\href{mailto:quanw@google.com}{quanw}\}@google.com
}

\begin{document}

\maketitle
\begin{abstract}
In this paper, we introduce a streaming keyphrase detection system that can be easily customized to accurately detect any phrase composed of words from a large vocabulary. The system is implemented with an end-to-end trained automatic speech recognition (ASR) model and a text-independent speaker verification model. To address the challenge of detecting these keyphrases under various noisy conditions, a speaker separation model is added to the feature frontend of the speaker verification model, and an adaptive noise cancellation (ANC) algorithm is included to exploit cross-microphone noise coherence. Our experiments show that the text-independent speaker verification model largely reduces the false triggering rate of the keyphrase detection, while the speaker separation model and adaptive noise cancellation largely reduce false rejections.
\end{abstract}

\section{Introduction}
In most voice assistive technologies, keyword spotting (\emph{a.k.a} wake word detection~\cite{kumatani2017direct}) is a common way to initiate the human-machine conversation (\emph{e.g.} ``OK Google'', ``Alexa'', or ``Hey Siri''). In recent years, keyword spotting techniques have evolved with many exciting advances, for example, using deep neural networks~\cite{chen2014small}, or end-to-end models~\cite{alvarez2019end,shan2018attention}.

However, most modern keyword spotting models are based on single or a few predefined phrases, often assuming the keyword is covered by a fixed-length window of audio. Supporting a new phrase usually requires re-training the entire system, which could be resource and time consuming.

In many scenarios, users would largely prefer a more seamless and natural interaction with the voice assistant without having to say a predefined keyword; especially for simple commands, such as ``Turn on the lights''. However, these interactions pose new challenges for conventional keyword spotting systems. In particular,
\begin{enumerate}[noitemsep,leftmargin=12pt]
    \item The system must be able to detect a large corpus of keyphrases.
    \item The keyphrases may have variable length, from single word (\emph{e.g.} ``Stop'') to longer sentences (\emph{e.g.} ``What is the weather tomorrow?''). The audio duration of the keyphrases could also vary depending on the speaker.
    \item The set of recognized keyphrases should be easily customizable without training and deploying new models.
\end{enumerate}

Instead of using a dedicated keyphrase detection model, we explore the possibility of using a generic ASR model that allows user-defined keyphrases, thereby providing greater flexibility to the users. A similar system was previously described in~\cite{he2017streaming}, where a Recurrent Neural Network Transducer (RNN-T) was trained to predict either phonemes or graphemes as subword units, thus allowing the detection of arbitrary keyphrases. However, a distinct challenge of a keyphrase detection that was not addressed in~\cite{he2017streaming} is being able to discriminate between the spoken keyphrases and noise in the background. This is especially difficult if the ambient noise includes speech that contains similar keyphrases. For example, a speaker on TV saying ``turn off the lights'' could easily false trigger the system.

Recognizing speech in a noisy, multi-talker environment, or the \textit{cocktail-party problem}, is an active area of research~\cite{simpsonCocktailParty, nachmani2020voice}. The human brain has the remarkable ability to identify and separate one person's voice from another~\cite{mcdermott2009cocktail}, especially if the speaker is familiar. One way the brain solves the cocktail-party problem is by using top-down attention to identify vocal features from a known speaker, while filtering out other irrelevant ambient sounds~\cite{pressnitzer2008perceptual}. In this paper, we represent vocal features of the enrolled speaker with neural network embeddings~\cite{wan2018generalized}, and use this information to suppress background speech from unknown speakers~\cite{Wang2020} in the feature frontend of the speaker verification model.

In addition, on devices where we have multiple microphones separated by a small distance (\emph{e.g.} smart home speakers), an adaptive noise cancellation algorithm can further enhance the speech signals by suppressing background noise.            

The original contributions of this paper include: (1) We adopt the state-of-the-art RNN-T model proposed in ~\cite{sainath2020e2e} and apply pruning~\cite{zhu2018prune} so that it can run continuously on device with significantly reduced CPU usage; (2) We combine the RNN-T-based ASR model with speaker verification and speaker separation models to achieve low false trigger and false rejection rates under various noise conditions; (3) We propose \textit{Speech Cleaner}, an adaptive noise cancellation algorithm that generalizes Hotword Cleaner~\cite{huang2019hotword} for generic speech recognition.

The rest of this paper is organized as follows. In Section~\ref{sec:system_overview}, we provide an overview of our keyphrase detection system, followed by detailed descriptions of the ASR model in Section~\ref{sec:asr_model}, speaker verification model in Section~\ref{sec:sv_model}, speaker separation model in Section~\ref{sec:vfl_model}, and adaptive noise cancellation in Section~\ref{sec:cleaner}. Two groups of experiments are presented in Section~\ref{sec:exp}. In Section~\ref{sec:exp_sv}, we demonstrate that a VoiceFilter-Lite model can largely reduce the Equal Error Rate (EER) of a standard text-independent speaker verification system under multi-talker scenarios. In Section~\ref{sec:exp_keyphrase}, we provide end-to-end evaluations of our keyphrase detection system under various noise conditions. Conclusions are drawn in Section~\ref{sec:conclusion}.

\section{Methods}
\label{sec:methods}
\begin{figure*}
	\centering
	\includegraphics[width=0.9\textwidth]{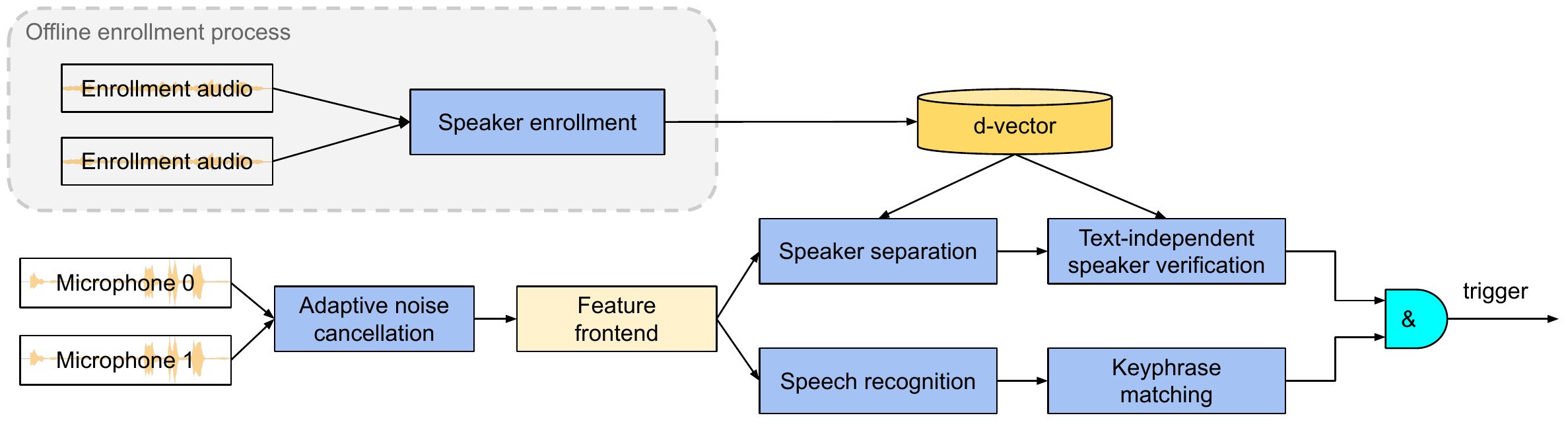}
	\caption{Diagram of the proposed keyphrase detection system. The d-vector is obtained in a separate offline enrollment process.}
	\label{fig:digram}
\end{figure*}

\subsection{System overview}
\label{sec:system_overview}

A diagram of the proposed keyphrase detection system is provided in Fig.~\ref{fig:digram}.

\subsubsection{Feature frontend} A shared feature frontend is used by all speech models in the system. This frontend first applies automatic gain control~\cite{prabhavalkar2015automatic} to the input audio, then extracts 32ms-long Hanning-windowed frames with a step of 10ms. For each frame, 128-dimensional log Mel-filterbank energies are computed in the range between 125Hz and 7500Hz. These filterbank energies are then stacked by 4 frames and subsampled by 3 frames, resulting in final features of 512 dimensions with a frame rate of 30ms.

\subsubsection{d-vector} The d-vector is an embedding vector that represents the voice characteristics of the enrolled user. It is obtained by prompting the user to follow an offline voice enrollment process~\cite{enrollmentblog,wang2020version}. At runtime, the d-vector is used in two ways: (1) It is used as a side input to the speaker separation model~\cite{Wang2020} to remove feature components not from the target speaker; (2) It represents the enrolled speaker in the speaker verification model.

\subsubsection{Keyphrase detection} The keyphrase detection system triggers only when both of the following conditions are met:
\begin{enumerate}[noitemsep,leftmargin=12pt]
    \item The text-independent speaker verification system successfully verifies the target enrolled user.
    \item The recognized text from the speech recognition model successfully matches one of the predefined keyphrases via regular expression pattern matching.
\end{enumerate}

Given this, there are two main sources of errors: (1) False accepts, where either a phrase other than the keyphrase or a keyphrase spoken by an unknown speaker (for example, in the background) triggers the detection system. (2) False rejects, where either the keyphrase was not recognized correctly by the ASR model, or the target user was mis-identified by the speaker verification system.

\subsection{Speech recognition}
\label{sec:asr_model}

The speech recognition model is an end-to-end RNN Transducer (RNN-T) model~\cite{graves2012sequence} with a similar architecture as proposed in ~\cite{he2019streaming, sainath2020e2e}. The target output vocabulary consists of 4096 word-pieces. The encoder network has 8 CIFG-LSTM layers~\cite{greff2016lstm} and the prediction network has 2 CIFG-LSTM layers. Each CIFG-LSTM layer has 2048 hidden units followed by a projection size of 640 units. The joint network has 640 hidden units and a softmax layer with 4096 units. To meet the memory constraints of running the speech recognition model continuously on-device, we shrink the model by applying 60\% sparsity~\cite{zhu2018prune} to each CIFG-LSTM layer in order to reduce the CPU usage, and consequently prolong the life of the device. The total model size is 42MB after sparsification and quantization~\cite{shangguan2019optimizing}. The model is trained on ~400K hours of multi-domain data including YouTube, voice search, farfield and telephony speech~\cite{narayanan2019longform}. We also add domain-ID to the model input during training and inference, which further improves the speech recognition quality in the target domain~\cite{sainath2020e2e}. 

In this work, we focus on home automation applications. To achieve this, we combine the domain-IDs for voice search and farfield domains during training, and use this ID during inference. However, since the target keyphrases tested in our work are common voice command queries, such as ``Stop'' or ``Turn on the light'', they appear frequently in the target domain training data. This in turn causes the ASR to have an implicit bias towards hypothesizing these keyphrases during inference.


\subsection{Speaker verification}
\label{sec:sv_model}

Many keyword spotting systems are shipped together with a speaker verification  (SV) model. The speaker verification model not only enables features such as personalized queries~\cite{multiuser} (\emph{e.g.} ``What's on my calendar?''), but also largely reduces the false accept rate of the keyword spotting system.

Since conventional keyword spotting systems only support single or a few keywords (\emph{e.g.} ``OK Google'' and ``Hey Google''), the speaker verification model shipped with them is also usually text-dependent. However, for a personalized keyphrase detection system that needs to support theoretically an infinite number of keyphrases, a text-independent speaker verification model must be used.

In this work, we use a text-independent model trained with the generalized end-to-end extended-set softmax loss~\cite{wan2018generalized,pelecanos2021dr}. Most of our training data are from a vendor collected multi-language speech query dataset covering 37 locales. We also added public datasets including LibriVox,
CN-Celeb~\cite{fan2020cn},
TIMIT~\cite{garofolo1993darpa},
and VCTK~\cite{yamagishi2019cstr}
to the training data for domain robustness. Multi-style training (MTR)~\cite{lippmann1987multi,ko2017study,kim2017generation} with SNR ranging from 3dB to 15dB is applied during the training process for noise robustness. The speaker verification model has 3 LSTM layers each with 768 nodes and a projection size of 256. The output of the last LSTM layer is then linearly transformed to the final 256-dimension d-vector. 

\subsection{Speaker separation}
\label{sec:vfl_model}


Since the ASR model is implicitly biased towards keyphrases in the target domain, we found that keyphrase detection has a low false rejection rate, even under noisy background conditions. In contrast, speaker verification systems are vulnerable to overlapping speech. For example, when the target user and an interfering speaker speak at the same time, the speaker verification system might reject the utterance, as the d-vector computed from overlapping speech would be very different to the d-vector derived from the target user speech alone.

Since speaker verification is critical to reducing false triggering, it is important to address the challenge of accurate speaker verification in multi-talker conditions. In this work, we use the VoiceFilter-Lite model~\cite{Wang2020} to enhance the input features from a single, enrolled speaker to the speaker verification model while masking out background speech.

Unlike other speech enhancement or separation models~\cite{hershey2016deep,kolbaek2017multitalker,Rao2019,Wang2019}, VoiceFilter-Lite has these benefits: (1) It directly enhances filterbank energies instead of the audio waveform, which largely reduces the number of runtime operations; (2) It supports streaming inference with low latency; (3) It uses an adaptive suppression strength, such that it is only effective on overlapping speech, avoiding unnecessary over-suppression; (4) It is optimized for on-device applications~\cite{shangguan2019optimizing}. For more details on the training data and model topology of VoiceFilter-Lite, please refer to~\cite{Wang2020}.

\subsection{Adaptive noise cancellation}
\label{sec:cleaner}

On devices which have more than one microphone such as smart speakers and mobile phones, an adaptive noise-cancellation (ANC) algorithm ~\cite{widrow1975adaptive} can be used to learn a filter that suppresses noise based on the correlation of the audio signals at multiple microphones during noise-only segments. 

Such an algorithm was proposed in ~\cite{huang2019hotword} for noise-robust keyword spotting. For our personalized keyphrase detection system, we use a module code-named \emph{Speech Cleaner}. Unlike Hotword Cleaner~\cite{huang2019hotword} where the adaptive filter coefficients are estimated using a FIFO buffer, in \emph{Speech Cleaner}, the adaptive filter coefficients are determined from a three second-long period of non-speech audio that precedes the speech signal. These coefficients are then kept frozen in order to suppress noise during the epoch containing speech.

\section{Experiments}
\label{sec:exp}

\begin{table}[t]
\centering
  \caption{Equal Error Rate (\%) of our speaker verification system under various noise conditions, with and without a VoiceFilter-Lite (VFL) model.}
  \label{tab:eer}
  \begin{tabular}{| c | c | c | c | c |}
    \hline
    \bf Noise & \bf \multirow{2}{*}{Room} & \bf SNR & \multicolumn{2}{c|}{\bf EER (\%)} \\ \cline{4-5}
    \bf source & & \bf (dB) & \bf No VFL & \bf With VFL \\  \hline
    \multicolumn{3}{|c|}{Clean} & 0.65 & 0.64 \\ \hline
    \multirow{6}{*}{Non-speech} & \multirow{3}{*}{Additive} & -5 & 5.30 & 5.23 \\ \cline{3-5}
    & & 0 & 2.04 & 2.01 \\ \cline{3-5}
    & & 5 & 1.22 & 1.22 \\ \cline{2-5}
     & \multirow{3}{*}{Reverb} & -5 & 6.51 & 6.53 \\ \cline{3-5}
    & & 0 & 2.90 & 2.91 \\ \cline{3-5}
    & & 5 & 1.60 & 1.59 \\ \hline
    \multirow{6}{*}{Speech} & \multirow{3}{*}{Additive} & -5 & 12.83 & \bf 4.24 \\ \cline{3-5}
    & & 0 & 8.34 & \bf 2.35 \\ \cline{3-5}
    & & 5 & 4.99 & \bf 1.47 \\ \cline{2-5}
     & \multirow{3}{*}{Reverb} & -5 & 17.76 & \bf 7.03 \\ \cline{3-5}
    & & 0 & 11.04 & \bf 3.63 \\ \cline{3-5}
    & & 5 & 6.41 & \bf 2.09 \\ \hline
  \end{tabular}
  \vspace{-0.3cm}
\end{table}

\subsection{Multi-talker speaker verification}
\label{sec:exp_sv}

Our first group of experiments focuses on addressing the multi-talker speaker verification challenge. We evaluate the standard speaker verification task under various noise conditions with and without a VoiceFilter-Lite model, while the noise source can be either non-speech noise or an interference speaker, and the room condition can be either additive or reverberant to simulate both near-field and far-field devices.

\subsubsection{Datasets}
For this evaluation, we use a vendor-provided English speech query dataset. There are 8,069 utterances from 1,434 speakers in the enrollment list, and 194,890 utterances from 1,241 speakers in the test list. The interference speech are from a separate English dev-set consisting of 220,092 utterances from 958 speakers. The non-speech noises are from various sources such as ambient noises recorded in silent environments, cafes, vehicles, and audio clips of music and sound effects downloaded from \href{https://www.gettyimages.com/}{gettyimages.com}.

\subsubsection{Results}
In Table~\ref{tab:eer}, we can see that under both clean and non-speech noise conditions, adding VoiceFilter-Lite does not affect the EER of the speaker verification system. This is expected because VoiceFilter-Lite uses an adaptive suppression strength, as explained in~\cite{Wang2020}. However, under speech noise conditions, VoiceFilter-Lite largely reduces the EER of the speaker verification system, under both additive and reverberant room conditions, and for various signal-to-noise ratio (SNR) setups. On average, VoiceFilter-Lite offers \textbf{a relative 67.4\% EER reduction} under speech noise conditions. Similar results had been reported by researchers using different models and data~\cite{Rao2019}.

\subsection{Keyphrase detection in the presence of ambient noise}
\label{sec:exp_keyphrase}

\begin{table}[t]
\centering
\caption{Overall end-to-end performance of our keyphrase detection system on a variety of datasets augmented with different noise sources. SV: speaker verification. VFL: VoiceFilter-Lite}
\label{tab:fr}
\begin{tabular}{|c|c|c|c|c|c|}
\hline
\multirow{2}{*}{\textbf{\begin{tabular}[c]{@{}c@{}}Noise\\ Source\end{tabular}}} & \multirow{2}{*}{\textbf{\begin{tabular}[c]{@{}c@{}}SNR\\ (dB)\end{tabular}}} & \multirow{2}{*}{\textbf{\begin{tabular}[c]{@{}c@{}}No \\ SV\end{tabular}}} & \multicolumn{3}{c|}{\textbf{With SV}} \\ \cline{4-6} 
 &  &  & \textbf{\begin{tabular}[c]{@{}c@{}}No\\ VFL\end{tabular}} & \textbf{\begin{tabular}[c]{@{}c@{}}VFL\\$\rightarrow$ASR\end{tabular}} & \textbf{\begin{tabular}[c]{@{}c@{}}VFL\\$\rightarrow$SV\end{tabular}} \\ \hline
\multicolumn{6}{|c|}{\multirow{2}{*}{\textbf{\begin{tabular}[c]{@{}c@{}}Synthetic Text-to-Speech generated keyphrases \\ False Rejection Rate (\%)\end{tabular}}}} \\
\multicolumn{6}{|c|}{} \\ \hline
\multicolumn{2}{|c|}{Clean} & 2.27 & 2.27 & 2.27 & 2.27 \\ \hline
\multirow{3}{*}{Non-speech} & -5 & 13.6 & 15.3 & 15.2 & \textbf{14.8} \\ \cline{2-6} 
 & 0 & 6.81 & 7.02 & 6.95 & \textbf{6.92} \\ \cline{2-6} 
 & 5 & 2.27 & 2.27 & 2.27 & 2.27 \\ \hline
\multirow{3}{*}{Speech} & -5 & 59.9 & 71.9 & 71.8 & \textbf{59.9} \\ \cline{2-6} 
 & 0 & 28.6 & 41.9 & 40.8 & \textbf{29.6} \\ \cline{2-6} 
 & 5 & 19.6 & 34.9 & 32.8 & \textbf{19.6} \\ \hline
\multicolumn{6}{|c|}{\multirow{2}{*}{\textbf{\begin{tabular}[c]{@{}c@{}}Vendor-provided keyphrases \\ False Rejection Rate (\%)\end{tabular}}}} \\
\multicolumn{6}{|c|}{} \\ \hline
\multicolumn{2}{|c|}{Clean} & 6.27 & 6.27 & 6.27 & 6.27 \\ \hline
\multirow{3}{*}{Non-speech} & -5 & 21.8 & 23.6 & 23.4 & 23.6 \\ \cline{2-6} 
 & 0 & 7.62 & 10.9 & 9.42 & 10.8 \\ \cline{2-6} 
 & 5 & 3.63 & 8.42 & 8.12 & 8.37 \\ \hline
\multirow{3}{*}{Speech} & -5 & 89.1 & 92.7 & 88.2 & \textbf{89.1} \\ \cline{2-6} 
 & 0 & 61.8 & 74.5 & 73.8 & \textbf{61.8} \\ \cline{2-6} 
 & 5 & 23.6 & 36.4 & 36.2 & \textbf{22.8} \\ \hline
\multicolumn{6}{|c|}{\multirow{2}{*}{\textbf{\begin{tabular}[c]{@{}c@{}}YouTube (no keyphrases)\\ False Acceptance / hour\end{tabular}}}} \\
\multicolumn{6}{|c|}{} \\ \hline
\multicolumn{2}{|c|}{} & 0.395 & 0.0354 & 0.0354 & \textbf{0.0286} \\ \hline
\end{tabular}
\vspace{-0.6cm}
\end{table}

Our second group of experiments focuses on evaluating the overall performance of the keyphrase detection system under various noise conditions. Specifically, we use two key metrics to evaluate the performance: (1) the number of false acceptance per hour (FA/h), which measures how many keyphrases are incorrectly accepted by the system; and (2) the false rejection rate (FRR), which measures the percentage of true keyphrases that are ignored by the system.

\subsubsection{Datasets}
To evaluate FA/h, we used a dataset consisting of 156 hours of English speech from curated and hand-annotated YouTube videos~\cite{narayanan2019longform}. This dataset is designed to mimic background noise, and contains no true keyphrases. As such, phrases in this dataset that trigger the detection system are considered false accepts. We generated d-vectors by enrolling human speakers from a custom, in-house database of speakers. 

To evaluate FRR, we used two datasets containing a set of commonly used keyphrases such as ``remind me to set an alarm", ``turn off the lights", and ``set a timer". First, we synthesized a dataset of 98 Text-to-Speech (TTS) speakers, each with 1000 keyphrases, using a previously published method~\cite{shen2018natural}. Additionally, we evaluated the performance on a set of vendor-provided keyphrases consisting of 61,555 utterances from 250 speakers with an average of 240 utterances per speaker. Each utterance was hand transcribed. For both datasets, we generated d-vectors from four enrollment utterances (e.g. ``Hey Google, remind me to water my plants"). Each speaker was enrolled separately to mimic single-user devices. The remaining utterances were used for evaluation. We augmented both datasets with either speech or non-speech background noise with reverberation at three different SNR levels using MTR.

To evaluate adaptive noise cancellation, we prepended each utterance with three seconds of silence before applying MTR. As a result, each utterance had three seconds of pure noise before the start of the main audio, which we used to estimate and freeze the \emph{Speech Cleaner} filter coefficients. We used the same noise sources and room configurations in all experiments.

\subsubsection{Results}

The overall performance (averaged across speakers) of our keyphrase detection system on the three above-mentioned datasets is shown in Table~\ref{tab:fr}. We observed that including speaker verification alone significantly decreased FA/h from 0.395 to 0.035 (rel. \textbf{91\%}). Adding a VoiceFilter-Lite (VFL) speaker separation model in the frontend of speaker verification, but not ASR, further reduced FA/h by improving speaker verification accuracy. Therefore, knowledge of speaker identity is sufficient to reduce false triggering.

This reduction in FA/h (no VFL), however, was accompanied by a 46.5\% increase in FRR in the multi-talker and a 3.08\% increase in the non-speech case (SNR = 0dB) relative to the model with no speaker verification. We observed a similar trend for the vendor-provided data with a rel. 20.6\% increase in FRR when speech background noise was added (SNR = 0dB). In both datasets, non-speech background noise resulted in far fewer false rejections than the multi-talker scenario. It is important to note that the increase in false rejections was primarily due to incorrect speaker verification (64\% of errors), rather than incorrect speech recognition. 

Adding a speaker separation (VFL) model to the feature frontend of speaker verification reduced the FRR from 41.9\% to 29.6\%, resulting in a 29.4\%  reduction in FRR in the SNR = 0dB multi-talker case relative to the model with only speaker verification. In particular, for both the TTS and vendor-provided datasets, adding speaker separation mitigated the increase in FRR caused by speaker verification. This improvement was due to the fact that VFL is effective at identifying and suppressing overlapping speech from a non-enrolled speaker, which in turn improved speaker verification accuracy. As a consequence of this all three models performed similarly in the non-speech background case. Notably, adding speaker separation to the feature frontend of the ASR alone did not produce a similar decrease in FRR, underscoring the fact that the false rejections in this keyphrase detection system are primarily due to speaker verification errors in the presence of speech background noise. 

Finally, to further improve the robustness of our keyphrase detection system to background noise, we included adaptive noise cancellation (ANC) in the feature frontends of both ASR and speaker verification. Relative to the model with only speaker verification, adding ANC reduced FRR by 68.3\% in the non-speech and 25.2\% in the speech background noise (SNR = 0dB) situations respectively. We refer the reader to Table~\ref{tab:anc} for a full description of the results.

Altogether, using both TTS and vendor-provided data, we have demonstrated that adding speaker verification, separation and adaptive noise cancellation results in a personalized keyphrase detection system that is robust to both background noise and overlapping speech.

\section{Conclusions}
\label{sec:conclusion}

\begin{table}[t]
\caption{False Rejection Rates (\%) of our keyphrase detection system on vendor-provided keyphrases with and without adaptive noise cancellation.}
\label{tab:anc}
\begin{tabular}{|c|c|c|c|c|c|}
\hline
\multirow{2}{*}{\textbf{\begin{tabular}[c]{@{}c@{}}Noise\\ Source\end{tabular}}} & \multirow{2}{*}{\textbf{\begin{tabular}[c]{@{}c@{}}SNR\\ (dB)\end{tabular}}} & \multicolumn{2}{c|}{\textbf{SV}} & \multicolumn{2}{c|}{\textbf{VFL $\rightarrow$ SV}} \\ \cline{3-6} 
 &  & \textbf{\begin{tabular}[c]{@{}c@{}}No \\ ANC\end{tabular}} & \textbf{\begin{tabular}[c]{@{}c@{}}With\\ ANC\end{tabular}} & \textbf{\begin{tabular}[c]{@{}c@{}}No \\ ANC\end{tabular}} & \textbf{\begin{tabular}[c]{@{}c@{}}With\\ ANC\end{tabular}} \\ \hline
\multicolumn{2}{|c|}{Clean} & 6.27 & 6.19 & 6.27 & 6.19 \\ \hline
\multirow{3}{*}{Non-speech} & -5 & 23.6 & 11.7 & 23.6 & \textbf{11.45} \\ \cline{2-6} 
 & 0 & 10.9 & 3.72 & 10.8 & \textbf{3.42} \\ \cline{2-6} 
 & 5 & 8.42 & 2.85 & 8.37 & \textbf{2.72} \\ \hline
\multirow{3}{*}{Speech} & -5 & 92.7 & 75.6 & 89.1 & \textbf{72.8} \\ \cline{2-6} 
 & 0 & 74.5 & 56.4 & 61.8 & \textbf{46.2} \\ \cline{2-6} 
 & 5 & 36.4 & 35.7 & 22.8 & \textbf{22.4} \\ \hline
\end{tabular}
\vspace{-0.4cm}
\end{table}

We proposed a streaming personalized keyphrase detection system that is highly robust to background noise and overlapping speech. We used a RNN-T based ambient ASR model that was pruned to fit on-device constraints and implicitly biased it towards voice commands via domain-id. To compensate for false triggering caused by biasing, we used a text-independent speaker verification model that rejected all keyphrases from non-enrolled speakers, which reduced FA/h by \textbf{91\%}. To mitigate the increased false rejections caused by speaker verification in the multi-talker scenario, we added a speaker separation model to the feature frontend of the speaker verification system. This resulted in a \textbf{67.4\%} reduction of speaker verification EER and a \textbf{29.4\%} reduction of FRR when the background contains overlapping speech. We also proposed \emph{Speech Cleaner}, a multi-microphone adaptive noise cancellation algorithm that further reduced FRR for noisy conditions.





\clearpage
\bibliographystyle{IEEEtran}
\bibliography{mybib}
\end{document}